\documentclass{article}
\usepackage{amsmath}
\usepackage{amsfonts}
\usepackage{amssymb}
\usepackage{graphicx}

\input{tcilatex}
\begin{document}

\title{Quantifying Rational Belief\thanks{%
Presented at MaxEnt 2009, the 29th International Workshop on Bayesian
Inference and Maximum Entropy Methods in Science and Engineering (July 5-10,
2009, Oxford, Mississippi, USA).}}
\author{Ariel Caticha \\
{\small Department of Physics, University at Albany-SUNY, }\\
{\small Albany, NY 12222, USA.}}
\date{}
\maketitle

\begin{abstract}
Some criticisms that have been raised against the Cox approach to
probability theory are addressed. Should we use a single real number to
measure a degree of rational belief? Can beliefs be compared? Are the Cox
axioms obvious? Are there counterexamples to Cox? Rather than justifying
Cox's choice of axioms we follow a different path and derive the sum and
product rules of probability theory as the unique (up to regraduations)
consistent representations of the Boolean \textsc{and} and \textsc{or}
operations.
\end{abstract}

\section{Introduction}

The objective of the Cox approach to probability theory is to develop tools
for reasoning under conditions of uncertainty \cite{Cox 46}\cite{Jaynes 03}%
\cite{Caticha 08}. The method proposed by Cox amounts to ranking statements
according to the extent to which one is rationally justified in believing
them. The ranking is implemented by associating to each statement a real
number meant to represent a degree of rational belief. It is perhaps
surprising that a bare minimum of rationality, namely, a requirement of
consistency, is sufficient to yield a precise \emph{quantitative} formalism.
Cox's remarkable theorem states that ranking according to degrees of
rational belief is equivalent to following the rules of probability theory.

The importance of Cox's approach is, first, that it allows one to represent
a partial state of knowledge as a consistent web of interconnected beliefs
and, second, that it solves the long standing problem of interpretation:
degrees of belief are to be manipulated according to the mathematical rules
of probability theory and therefore no mistakes will ever be made if we call
them \textquotedblleft probabilities\textquotedblright . These are not
modest claims and it is only appropriate that the Cox approach be subjected
to a severe critical scrutiny. The purpose of this paper is to address some
of the criticisms that have been raised over the years.

A thoughtful overview and general criticism of induction theories appears in
the work of J. D. Norton \cite{Norton 07}. He points out that in order to
accept the Cox argument one must be convinced that beliefs come in numerical
degrees, that beliefs can be compared, and furthermore, that they must be
transitive (if $a$ is preferred over $b$, and $b$ over $c$, then $a$ is
preferred over $c$). Not obvious at all, he claims, a single number may not
be sufficient to capture the richness of our beliefs which could very well
be intransitive or even incommensurate. And the doubts do not end there.

Cox assumes as one of his axioms that the degree of belief in a proposition $%
a$ assuming that $b$ is true, which we write as $[a|b]$, is rigidly related
to the degree corresponding to its negation, $[\tilde{a}|b]$, through some
definite but initially unspecified function $f$, 
\begin{equation}
\lbrack \tilde{a}|b]=f\left( [a|b]\right) \text{~.}  \label{negation f}
\end{equation}%
To some of us this is intuitively reasonable: the more one believes in $a|b$%
, the less one believes in $\tilde{a}|b$. Not obvious, writes Norton,
\textquotedblleft if one does not prejudge what `belief'\ must be, the
assumption of this specific functional dependency is alien and
arbitrary\textquotedblright .\ 

A second Cox axiom is that the degree of belief of \textquotedblleft $a$ 
\textsc{and} $b$ assuming $c$,\textquotedblright\ written as $[ab|c]$, must
depend on $[a|c]$ and $[b|ac]$, 
\begin{equation}
\lbrack ab|c]=g\left( [a|c],[b|ac]\right) ~.  \label{conjunction g}
\end{equation}%
This is also very reasonable. When asked to check whether \textquotedblleft $%
a$ \textsc{and} $b$\textquotedblright\ is true, we first look at $a$; if $a$
turns out to be false the conjunction is false and we need not bother with $%
b $; therefore $[ab|c]$ must depend on $[a|c]$. If $a$ turns out to be true
we need to take a further look at $b$; therefore $[ab|c]$ must also depend
on $[b|ac]$. Norton objects that it may be reasonable, but it is also an
assumption \textquotedblleft likely to be uncontroversial only for someone
who already believes that plausibilities are probabilities and has tacitly
in mind that we must eventually recover the product rule\textquotedblright\ 
\cite{Norton 07}.

Strictly $[ab|c]$ could in principle depend on all four quantities $[a|c]$, $%
[b|c]$, $[a|bc]$ and $[b|ac]$, an objection that has a long history. It was
partially addressed by Tribus \cite{Tribus 69} and then by Smith and
Erickson \cite{Smith Erickson 90} but, unfortunately, as has been
convincingly pointed out by Garrett \cite{Garrett 96}, their arguments are
not completely satisfactory.

Yet another objection has been raised by Halpern \cite{Halpern 99}. He shows
that in finite domains it is possible to satisfy the consistency constraint
that follows from the associativity of the Boolean \textsc{and}, $%
(ab)c=a(bc) $, without requiring that the function $g$ in eq.(\ref%
{conjunction g}) be itself associative. This allows him to construct
counterexamples to Cox's theorem.

In section 2 we discuss degrees of rational belief and why we are justified
in representing them by real numbers. Then to (partially) counter the
objection that the Cox axioms are intuitive only to those who are already
convinced of the results we reformulate the Cox theory in terms of axioms
that differ from the usual ones. The idea is to construct a representation
of the Boolean \textsc{and} and \textsc{or} by focusing on their associative
and distributive properties rather than on the operation of negation. We
then argue that it is the nature of our goal --- to construct an inductive
logic of \emph{general} applicability --- that allows us to escape Halpern's
criticism, and also to give a proper treatment to the Tribus-Smith-Erickson
objection. In section 3 an associativity constraint is used to derive the
sum rule rather than the product rule as Cox had originally done, and in
section 4 we focus on the distributive property of \textsc{and} over \textsc{%
or} to obtain the product rule. The negation function $f$ in eq.(\ref%
{negation f}) and the functional equation associated with it are completely
avoided.

Our subject is degrees of rational belief but the algebraic approach
followed here can be pursued in its own right irrespective of any
interpretation. It was used in \cite{Caticha 98} to derive the manipulation
rules for complex numbers interpreted as quantum mechanical amplitudes. It
was also used by K. Knuth \cite{Knuth 03} in the purely mathematical problem
of assigning real numbers (valuations) on general distributive lattices.

\section{Degrees of rational belief}

Different individuals may hold different beliefs and it is certainly
important to figure out what those beliefs might be --- perhaps by observing
their gambling behavior --- but this is not our present concern. Our
objective is neither to assess nor to describe the subjective beliefs of any
particular individual. Instead we deal with the altogether different but
very common problem that arises when we are confused and we want some
guidance about what we are \emph{supposed }to believe. Our concern here is
not so much with beliefs as they actually are, but rather, with beliefs as
they ought to be.

Rational beliefs are constrained beliefs. Indeed, \emph{the essence of
rationality lies precisely in the existence of some constraints}. The
problem, of course, is to figure out what those constraints might be. We
need to identify \emph{normative criteria of rationality}. It must be
stressed that the beliefs discussed here are meant to be those held by an
idealized rational individual who is not subject to practical human
limitations. We are concerned with those ideal standards of rationality that
we ought to strive to attain at least when discussing scientific\ matters.

Here is our first criterion of rationality: whatever guidelines we pick they
must be of general applicability---otherwise they fail when most needed,
namely, when not much is known about a problem. Different rational
individuals can reason about different topics, or about the same subject but
on the basis of different information, and therefore they could hold
different beliefs, but they must agree to follow the same rules.

As a second criterion of (extremely idealized) rationality we require
theories that allow quantitative reasoning. The obvious question concerns
the type of quantity that will represent the intensity of beliefs. Discrete
categorical variables are not adequate for a theory of general
applicability; we need a much more refined scheme.

Do we believe statement $a$ more or less than statement $b$? Are we even
justified in comparing statements $a$ and $b$? The problem with statements
is not that they cannot be compared but rather that the comparison can be
carried out in too many different ways. We can classify statements according
to the degree we believe they are true, their plausibility; or according to
the degree that we desire them to be true, their utility; or according to
the degree that they happen to bear on a particular issue at hand, their
relevance. We can even compare propositions with respect to the minimal
number of bits that are required to state them, the description length. The
detailed nature of our relations to statements is too complex to be captured
by a single real number. What we claim is that a single real number is
sufficient to measure one specific feature, the sheer intensity of rational
belief. This should not be too controversial because it amounts to a
tautology: an \textquotedblleft intensity\textquotedblright\ is precisely
the type of quantity that admits no more qualifications than that of being
more intense or less intense; it is captured by a single real number.

However, some preconception about our subject is unavoidable; we need some
rough notion that a belief is not the same thing as a desire. But how can we
know that we have captured pure belief and not belief contaminated with some
hidden desire? Strictly we can't. We hope that our mathematical description
captures a sufficiently purified notion of rational belief, and we can claim
success only to the extent that the formalism proves to be useful. Since
some preconceived notions are needed, here is one: we take it to be a
defining feature of the intensity of \emph{rational }beliefs that if $a$ is
more believable than $b$, and $b$ more than $c$, then $a$ is more believable
than $c$. Such transitive rankings can be implemented using real numbers we
are again led to claim that degrees of rational belief can be represented by
real numbers.

The notation we use is fairly standard: given any two statements $a$ and $b$
the disjunction \textquotedblleft $a$ \textsc{or} $b$\textquotedblright\ and
the conjunction \textquotedblleft $a$ \textsc{and} $b$\textquotedblright\
are denoted respectively by $a\vee b$ and $ab$. Typically we want to
quantify our beliefs in $a\vee b$ and in $ab$ in the context of some
background information $c$, which we write as $ab|c$. The real number that
represents the degree of belief in $a|b$ will initially be denoted by $[a|b]$%
. Degrees of rational belief will range from the extreme of total certainty, 
$[a|a]=v_{T}$, to total disbelief, $[\tilde{a}|a]=v_{F}$. Note that the
transitivity of the ranking scheme implies that there is a single value $%
v_{F}$ and a single $v_{T}$.

Here is a second preconceived notion: in order to be rational our beliefs in 
$a\vee b$ and $ab$ must be somehow related to our separate beliefs in $a$
and $b$. Since the goal is to design a quantitative theory, we require that
these relations be represented by some functions $F$ and $G$,%
\begin{equation}
\lbrack a\vee b|c]=F([a|c],[b|c],[a|bc],[b|ac])  \label{OR F}
\end{equation}%
and 
\begin{equation}
\lbrack ab|c]=G([a|c],[b|c],[a|bc],[b|ac])~.  \label{AND G}
\end{equation}%
Note the \emph{qualitative} nature of this assumption: what is being
asserted is the existence of some unspecified functions $F$ and $G$ and not
their specific functional forms. The same $F$ and $G$ are meant to apply to
all propositions; what is being \emph{designed} is a single inductive scheme
of universal applicability. Note further that unlike eq.(\ref{conjunction g}%
) the arguments of $F$ and $G$ include all four possible degrees of belief
in $a$ and $b$ in the context of $c$ and not any potentially questionable
subset.

Since it is quite inconceivable that a smooth change in, say, $[a|c]$ could
lead to anything but a smooth change in $[a\vee b|c]$ and $[ab|c]$, we will
assume that the functions $F$ and $G$ are sufficiently smooth and well
behaved. Indeed, should it turn out that $F$ and $G$ require kinks or
discontinuities we would probably feel justified in throwing the whole
scheme away. (However, smoothness might not be necessary. See \cite{Aczel 66}%
.)

Our method is one of eliminative induction: now that we have identified a
sufficiently broad class of theories---quantitative theories of universal
applicability, with degrees of belief represented by real numbers and the
operations of conjunction and disjunction represented by functions---we can
start weeding the unacceptable ones out.

\section{The sum rule}

We start with the function $F$ that represents \textsc{or}. The space of
functions of four arguments is very large. To narrow down the field we
initially restrict ourselves to propositions $a$ and $b$ that are mutually
exclusive in the context of $d$. Thus, 
\begin{equation}
\lbrack a\vee b|d]=F([a|d],[b|d],v_{F},v_{F})~,
\end{equation}%
which effectively restricts $F$ to a function of only two arguments.

\noindent \textbf{The associativity constraint:}

We require that the assignment of degrees of belief be consistent---if a
degree of belief can be computed in two different ways the two ways must
agree---how else could we claim to be rational? All functions $F$ that fail
to satisfy this constraint must be discarded.

Consider any three mutually exclusive statements $a$, $b$, and $c$ in the
context of a fourth $d$. The consistency constraint that follows from the
associativity of the Boolean \textsc{or}, $(a\vee b)\vee c=a\vee (b\vee c)$,
is remarkably constraining. It essentially determines the function $F$.
Start from%
\begin{equation}
\lbrack a\vee b\vee c|d]=F\left( [a\vee b|d],[c|d]\right) =F\left(
[a|d],[b\vee c|d]\right) ~,
\end{equation}%
and using $F$ once again for $[a\vee b|d]$ and for $[b\vee c|d]$, we get 
\begin{equation}
F\{F\left( [a|d],[b|d]\right) ,[c|d]\}=F\{[a|d],F\left( [b|d],[c|d]\right) \}
\end{equation}%
If we call $[a|d]=x$, $[b|d]=y$, and $[c|d]=z$, then 
\begin{equation}
F\{F(x,y),z\}=F\{x,F(y,z)\}  \label{assoc constraints}
\end{equation}%
The function $F$ must obey (\ref{assoc constraints}) for arbitrary choices
of the propositions $a$, $b$, $c$, and $d$.

Halpern has raised the following objection \cite{Halpern 99}. Suppose we
have a belief function $[a|d]$ that associates a real number to each pair of
propositions $a$ and $d$. He observes that if the total number of such
propositions is finite (a discrete universe of discourse) then the triples $%
(x,y,z)$ to be used in (\ref{assoc constraints}) do not form a dense set and
therefore we are not allowed to conclude that the function $F$ must itself
be associative for arbitrary values of $x$, $y$, and $z$. Thus, in finite
universes of discourse it is possible to design models of inference that are
consistent without being equivalent to probability theory, and Halpern
constructs an explicit example.

The reply to Halpern's objection is not to be found in any flaw in his
mathematics. We must rather focus on the larger project at hand. We are
concerned with designing a theory of inference of \emph{universal }%
applicability, a single scheme to be used by all rational individuals
irrespective of their state of knowledge or of subject matter. One
individual might assign a plausibility $[a|d]=x$ while another, who is in
possession of different information, might assign $[a|d]^{\prime }=x^{\prime
}$, while a third would assign $x^{\prime \prime }$ and so on. Thus the
values $x$ form a dense set, not because the allowed propositions $a$, $b$%
,... are themselves dense, but rather because the belief functions $[\cdot
|\cdot ]$, $[\cdot |\cdot ]^{\prime }$,... are dense. Furthermore, the same
general purpose scheme must be applicable to arbitrary subject matter, not
just to one particular discrete set, but also to continuous sets of
propositions. We conclude that \emph{in order to be of universal
applicability} the function $F$ must indeed be associative and satisfy (\ref%
{assoc constraints}) for arbitrary values of $(x,y,z)$.

\noindent \textbf{The general solution and its regraduation:}

By straightforward substitution one can check that eq.(\ref{assoc
constraints}) is satisfied if 
\begin{equation}
F\left( x,y\right) =\phi ^{-1}\left( \phi \left( x\right) +\phi \left(
y\right) \right) \,,  \label{xi1}
\end{equation}%
where $\phi $ is an arbitrary invertible function. It has been shown that
this is also the \emph{general} solution \cite{Cox 46}\cite{Aczel 66}. Given 
$\phi $ one can calculate $F$ and, conversely, given $F$ one can calculate
the corresponding $\phi $. Eq.(\ref{xi1}) can be rewritten as 
\begin{equation}
\phi \left( F\left( x,y\right) \right) =\phi \left( x\right) +\phi \left(
y\right) \quad \text{or}\quad \phi \left( \lbrack a\vee b|d]\right) =\phi
\left( \lbrack a|d]\right) +\phi \left( \lbrack b|d]\right) ~.
\end{equation}%
This last form is the pivotal point of the whole argument: it shows that
instead of representing degrees of belief along the scale provided by the
numbers $[a|d]$, we can equally well regraduate to a new scale given by $\xi
\left( a|d\right) =\phi \left( \lbrack a|d]\right) $. The original and the
regraduated scales are equivalent because being invertible the function $%
\phi $ is monotonic and preserves the ranking of propositions. However, the
regraduated scale is much more convenient because the \textsc{or} operation
is now represented by a much simpler sum rule, \ 
\begin{equation}
\xi \left( a\vee b|d\right) =\xi \left( a|d\right) +\xi \left( b|d\right) ~.
\label{sumxi}
\end{equation}

The regraduated $\xi _{F}=\phi (\nu _{F})$ is easy to evaluate. Setting $d=%
\tilde{a}$ in eq.(\ref{sumxi}) gives $\xi \left( a\vee b|\tilde{a}\right)
=\xi \left( a|\tilde{a}\right) +\xi \left( b|\tilde{a}\right) $. Since $%
a\vee b|\tilde{a}$ is true if and only if $b|\tilde{a}$ is true, the
corresponding degrees of belief must coincide, $\xi \left( a\vee b|\tilde{a}%
\right) =\xi \left( b|\tilde{a}\right) $, and therefore $\xi \left( a|\tilde{%
a}\right) =\xi _{F}=0$.

\noindent \textbf{The general sum rule:}

The restriction to mutually exclusive propositions in the sum rule eq.(\ref%
{sumxi}) can easily be lifted. Any proposition $a$ can be written as the
disjunction of two mutually exclusive ones, $a=(ab)\vee (a\tilde{b})$ and
similarly $b=(ab)\vee (\tilde{a}b)$. Therefore for any two \emph{arbitrary}
propositions $a$ and $b$ we have 
\begin{equation}
a\vee b=(ab)\vee (a\tilde{b})\vee (\tilde{a}b)
\end{equation}%
Since each of the terms on the right are mutually exclusive the sum rule (%
\ref{sumxi}) applies, 
\begin{eqnarray}
\xi (a\vee b|d) &=&\xi (ab|d)+\xi (a\tilde{b}|d)+\xi (\tilde{a}b|d)+[\xi
(ab|d)-\xi (ab|d)]  \notag \\
&=&\xi (ab\vee a\tilde{b}|d)+\xi (ab\vee \tilde{a}b|d)-\xi (ab|d)~,
\end{eqnarray}%
which leads to the general sum rule, 
\begin{equation}
\xi (a\vee b|d)=\xi (a|d)+\xi (b|d)-\xi (ab|d)~.  \label{sum rule}
\end{equation}

\section{The product rule}

Next we consider the function $G$ that represents \textsc{and}. The space of
functions of four arguments is very large so we first narrow it down to just
two. Then, we impose a consistency constraint that follows from the
distributive properties of the Boolean \textsc{and} and \textsc{or.} A
trivial regraduation yields the product rule of probability theory.

\noindent \textbf{From four arguments down to two:}

We will separately consider special cases where the function $G$ depends on
only two arguments, then three, and finally all four arguments. Using
commutivity, $ab=ba$, the possibilities are seven: 
\begin{eqnarray}
\xi (ab|c) &=&G^{(1)}[\xi (a|c),\xi (b|c)] \\
\xi (ab|c) &=&G^{(2)}[\xi (a|c),\xi (a|bc)] \\
\xi (ab|c) &=&G^{(3)}[\xi (a|c),\xi (b|ac)] \\
\xi (ab|c) &=&G^{(4)}[\xi (a|bc),\xi (b|ac)] \\
\xi (ab|c) &=&G^{(5)}[\xi (a|c),\xi (b|c),\xi (a|bc)] \\
\xi (ab|c) &=&G^{(6)}[\xi (a|c),\xi (a|bc),\xi (b|ac)] \\
\xi (ab|c) &=&G^{(7)}[\xi (a|c),\xi (b|c),\xi (a|bc),\xi (b|ac)]
\end{eqnarray}%
Since the method aims at general applicability the arguments of $%
G^{(1)}\ldots G^{(7)}$ can be varied independently.

First some notation: complete certainty is denoted $\xi _{T}$, \ while
complete disbelief is $\xi _{F}=0$. Derivatives are denoted with a
subscript: the derivative of $G^{(3)}(x,y)$ with respect to its second
argument $y$ is $G_{2}^{(3)}(x,y)$.

\noindent \textbf{Type 1: }$\xi (ab|c)=G^{(1)}[\xi (a|c),\xi (b|c)]$. The
function $G^{(1)}$ is unsatisfactory because it does not take possible
correlations between $a$ and $b$ into account. For example, when $a$ and $b$
are mutually exclusive --- for example $b=\tilde{a}d$, for some arbitrary $d$
--- $\xi (ab|c)=\xi _{F}$ but there are no constraints on either $\xi
(a|c)=x $ or $\xi (b|c)=y$. Thus, in order that $G^{(1)}(x,y)=\xi _{F}$ for
arbitrary choices of $x$ and $y$, $G^{(1)}$ must be a constant which is
unacceptable.

\noindent \textbf{Type 2: }$\xi (ab|c)=G^{(2)}[\xi (a|c),\xi (a|bc)]$. This
function is unsatisfactory because it overlooks the plausibility of $b|c$ 
\cite{Smith Erickson 90}. For example: let $a=$ \textquotedblleft $X$ is
big\textquotedblright\ and $b=$ \textquotedblleft $X$ is big and
green\textquotedblright\ so that $ab=b$. Then 
\begin{equation}
\xi (b|c)=G^{(2)}[\xi (a|c),\xi (a|abc)]\quad \text{or}\quad \xi
(b|c)=G^{(2)}[\xi (a|c),\xi _{T}]~,
\end{equation}%
which is clearly unsatisfactory since \textquotedblleft
green\textquotedblright\ does not figure anywhere on the right hand side.

\noindent \textbf{Type 3: }$\xi (ab|c)=G^{(3)}[\xi (a|c),\xi (b|ac)]$. This
function turns out to be satisfactory.

\noindent \textbf{Type 4: }$\xi (ab|c)=G^{(4)}[\xi (a|bc),\xi (b|ac)]$. This
function strongly violates common sense: when $a=b$ we have $\xi
(a|c)=G^{(4)}(\xi _{T},\xi _{T})$, so that $\xi (a|c)$ takes the same
constant value irrespective of what $a$ might be \cite{Smith Erickson 90}.

\noindent \textbf{Type 5: }$\xi (ab|c)=G^{(5)}[\xi (a|c),\xi (b|c),\xi
(a|bc)]$. This function turns out to be equivalent either to $G^{(1)}$ or to 
$G^{(3)}$ and can therefore be ignored. The proof follows from
associativity, $(ab)c|d=a(bc)|d$, which leads to the constraint 
\begin{eqnarray*}
&&G^{(5)}\left[ G^{(5)}[\xi (a|d),\xi (b|d),\xi (a|bd)],\xi
(c|d),G^{(5)}[\xi (a|cd),\xi (b|cd),\xi (a|bcd)]\right] \\
&=&G^{(5)}[\xi (a|d),G^{(5)}[\xi (b|d),\xi (c|d),\xi (b|cd)],\xi (a|bcd)]
\end{eqnarray*}%
and, with the appropriate identifications, 
\begin{equation}
G^{(5)}[G^{(5)}(x,y,z),u,G^{(5)}(v,w,s)]=G^{(5)}[x,G^{(5)}(y,u,w),s]~.
\end{equation}%
Since the variables $x,y\ldots s$ can be varied independently of each other
we can take a partial derivative with respect to $z$, 
\begin{equation}
G_{1}^{(5)}[G^{(5)}(x,y,z),u,G^{(5)}(v,w,s)]G_{3}^{(5)}(x,y,z)=0~.
\end{equation}%
Therefore, either 
\begin{equation}
G_{3}^{(5)}(x,y,z)=0\quad \text{or}\quad
G_{1}^{(5)}[G^{(5)}(x,y,z),u,G^{(5)}(v,w,s)]=0~.
\end{equation}%
The first possibility says that $G^{(5)}$ is independent of its third
argument which means that it is of the type $G^{(1)}$ that has already been
ruled out. The second possibility says that $G^{(5)}$ is independent of its
first argument which means that it is already included among the type $%
G^{(3)}$.

\noindent \textbf{Type 6: }$\xi (ab|c)=G^{(6)}[\xi (a|c),\xi (a|bc),\xi
(b|ac)]$. This function turns out to be equivalent either to $G^{(3)}$ or to 
$G^{(4)}$ and can therefore be ignored. The proof---which we omit because it
is identical to the proof above for type 5---also follows from
associativity, $(ab)c|d=a(bc)|d$.

\noindent \textbf{Type 7: }$\xi (ab|c)=G^{(7)}[\xi (a|c),\xi (b|c),\xi
(a|bc),\xi (b|ac)]$. This function turns out to be equivalent either to $%
G^{(5)}$ or $G^{(6)}$ and can therefore be ignored. Again the proof which
uses associativity, $(ab)c|d=a(bc)|d$, is omitted because it is identical to
type 5.

We conclude that the possible functions $G$ that are viable candidates for a
general theory of inductive inference are equivalent to type $G^{(3)}$, 
\begin{equation}
\xi (ab|c)=G[\xi (a|c),\xi (b|ac)]~.  \label{G xi}
\end{equation}

\noindent \textbf{The distributivity constraint:}

Consider three statements $a$, $b$, and $c$, where the last two are mutually
exclusive, in the context of a fourth, $d$. Distributivity, $a\left( b\vee
c\right) =ab\vee ac$, implies that $\xi \left( a\left( b\vee c\right)
|d\right) $ can be computed in two ways, $\xi \left( a\left( b\vee c\right)
|d\right) =\xi \left( \left( ab|d\right) \vee \left( ac|d\right) \right) $.
Using eq.(\ref{sumxi}) and (\ref{G xi}) leads to 
\begin{equation}
G\left( \xi \left( a|d\right) ,\xi \left( b|ad\right) +\xi \left(
c|ad\right) \right) =\text{ }G[\xi \left( a|d\right) ,\xi \left( b|ad\right)
]+G[\xi \left( a|d\right) ,\xi \left( c|ad\right) ]~.
\end{equation}%
Therefore, the requirement of distributivity constrains $G$ to satisfy%
\begin{equation}
G\left( u,v+w\right) =G\left( u,v\right) +G\left( u,w\right) ~,
\label{dist R}
\end{equation}%
\newline
where $\xi \left( a|d\right) =u$, $\xi \left( b|ad\right) =v$, and $\xi
\left( c|ad\right) =w$. To solve this constraint let $v+w=z$.
Differentiating with respect to $v$ and $w$ gives $\partial ^{2}\,G\left(
u,z\right) /\partial z^{2}=0$, so that $G$ is linear in its second argument, 
$G(u,v)=A(u)v+B(u)$. Substituting back into eq.(\ref{dist R}) gives $B(u)=0$.

To determine the function $A(u)$ \cite{N Caticha 09} we note that the degree
to which we believe in $ad|d$ is exactly the degree to which we believe in $%
a|d$ by itself. Therefore,  
\begin{equation}
\xi (a|d)=\xi (ad|d)=G[\xi (a|d),\xi (d|ad)]=G[\xi (a|d),\xi _{T}]\quad 
\text{or}\quad u=A(u)\xi _{T}\,,
\end{equation}%
which means that 
\begin{equation}
G\left( u,v\right) =uv/\xi _{T}\quad \text{or\quad }\xi \left( ab|d\right)
=\xi \left( a|d\right) \xi \left( b|ad\right) /\xi _{T}\,.
\end{equation}%
The constant $\xi _{T}$ is easily regraduated away: just normalize $\xi $ to 
$p=\xi /\xi _{T}$. In the regraduated scale the \textsc{and} operation is
represented by a simple product rule, \ 
\begin{equation}
p\left( ab|d\right) =p\left( a|d\right) \,p\left( b|ad\right) ~,
\label{p rule}
\end{equation}%
\newline
while the sum rule, eq.(\ref{sum rule}), remains unaffected, 
\begin{equation}
p\left( a\vee b|d\right) =p\left( a|d\right) +p\left( b|d\right) -p(ab|d).
\label{s rule}
\end{equation}%
\newline
Degrees of belief $p$ measured in this particularly convenient regraduated
scale can be called \textquotedblleft probabilities\textquotedblright .\ The
degrees of belief $\xi $ range from total disbelief $\xi _{F}=0$ to total
certainty $\xi _{T}$. The corresponding regraduated values are $p_{F}=0$ and 
$p_{T}=1$.

\section{Conclusion}

Probability theory is the unique method of rational, quantitative and
consistent inductive inference that can claim to be of general
applicability. It focuses on degrees of rational belief and not on other
qualities such as simplicity, explanatory power, degree of confirmation,
desirability, or amount of information. The reason the method is unique\ is
not because we have succeeded in formulating a precise and rigorous
definition of rationality. Rather, the method is unique for the more modest
reason that it is the only one left after obvious irrationalities --- such
as inconsistencies --- have been weeded out.

\noindent \textbf{Acknowledgements: }I am grateful to K. Knuth, and C. Rodr%
\'{\i}guez for many discussions on these topics, and especially to Nestor
Caticha who drew my attention to an error and indicated the appropriate
correction. I also thank T. Loredo who, after completion of this paper, drew
my attention to reference \cite{Van Horn 03} which also invokes the
requirement of universality to evade Halpern's objection.


\begin{thebibliography}{99}
\bibitem{Cox 46} R. T. Cox, Am. J. Phys. \textbf{14}, 1-13 (1946).

\bibitem{Jaynes 03} E. T. Jaynes, \emph{Probability Theory: The Logic of
Science}, ed. by L. Bretthorst (Cambridge U.P., 2003).

\bibitem{Caticha 08} A. Caticha, \emph{Lectures on Probability, Entropy, and
Statistical Physics} (MaxEn08, S\~{a}o Paulo, 2008)
(arXiv.org/abs/0808.0012).

\bibitem{Norton 07} J. D. Norton, Brit. J. Phil. Sci. \textbf{58}, 141
(2007).

\bibitem{Tribus 69} M. Tribus, \emph{Rational Descriptions, Decisions and
Designs }(Pergamon, 1969).

\bibitem{Smith Erickson 90} C. R. Smith, G. J. Erickson, p.17 in \emph{%
Maximum Entropy and Bayesian Methods} ed. by P. F. Foug\`{e}re (Kluwer,
1990).

\bibitem{Garrett 96} A. Garrett, p.175 in \emph{Maximum Entropy and Bayesian
Methods} ed. by G. R. Heidbreder (Kluwer, 1996).

\bibitem{Halpern 99} J. Y. Halpern, Journal of Artificial Intelligence
Research \textbf{10}, 67 (1999).

\bibitem{Caticha 98} A. Caticha, Phys. Rev. \textbf{A57}, 1572 (1998)
(arXiv.org/abs/quant-ph/9804012).

\bibitem{Knuth 03} K. H. Knuth, p. 204\ in \emph{Bayesian Inference and
Maximum Entropy Methods in Science and Engineering}, ed. by G.J. Erickson
and Y. Zhai, AIP Conf. Proc. \textbf{707} (2003).

\bibitem{Aczel 66} J. Acz\'{e}l, \emph{Lectures on Functional Equations and
Their Applications} (Academic Press, New York, 1996).

\bibitem{Van Horn 03} K. Van Horn, Int. J. Approx. Reasoning \textbf{34}, 3
(2003).

\bibitem{N Caticha 09} This argument is due to N. Caticha, private
communication (2009).
\end{thebibliography}
\end{document}